\documentclass[amsmath,amssymb,aps,reprint,prb]{revtex4-1}

\usepackage{chemformula} 
\usepackage[T1]{fontenc} 

\usepackage{graphicx}

\begin{document} 

\author{Alexis Fiset-Cyr}
\affiliation{Department of Physics, University of Sherbrooke, Sherbrooke, Quebec, Canada, J1K 2R1.}
\author{Dan Dalacu}
\email{dan.dalacu@nrc.ca}
\author{Sofiane Haffouz}
\author{Philip J. Poole}
\author{Jean Lapointe}
\author{Geof C. Aers}
\author{Robin L. Williams}

\affiliation{National Research Council of Canada, Ottawa, Ontario, Canada, K1A 0R6.}

\title{\textit{In-situ} tuning of individual position-controlled nanowire quantum dots via laser-induced intermixing}



\preprint{APS/123-QED}




\begin{abstract}

We demonstrate an $in$-$situ$ technique to tune the emission energy of semiconductor quantum dots. The technique is based on laser-induced atomic intermixing applied to nanowire quantum dots grown using a site-selective process that allows for the deterministic tuning of individual emitters.  A tuning range of up to 15\,meV is obtained with a precision limited by the laser exposure time. A distinct saturation of the energy shift is observed, which suggests an intermixing mechanism relying on grown-in defects that are subsequently removed from the semiconductor material during annealing. The ability to tune different emitters into resonance with each other will be required for fabricating remote quantum dot-based sources of indistinguishable photons for secure quantum networks.

\end{abstract}

\maketitle 
\newpage

A prime motivator for the development of single photon sources is that of secure optical communications based on quantum key distribution (QKD). Implementations such as quantum repeater-based\cite{Lo_Science1999} entanglement distribution and measurement-device-independent QKD\cite{Lo_PRL2012} require the generated photons to be indistinguishable one from the other\cite{Santori_NAT2002} (i.e. perfect overlap of the photon wave-packets in all degrees of freedom: energy, time, space, polarization). Quantum dot-based sources are a promising candidate for implementation in QKD networks: they can be extremely efficient if incorporated within appropriate photonic environments i.e. cavities\cite{Yoshie_APL2001,Moreau_APL2001}, waveguides\cite{Bulgarini_NL2014}, they have demonstrated high purity on-demand single photon emission\cite{Aharonovich_NP2016} and sequentially emitted photons from the same dot can be highly indistinguishable when using resonant excitation\cite{Loredo_Opt2016,Wang_PRL2016}. 

For use in distributed QKD networks however, indistinguishable photons from separate quantum dots are required and this has proven challenging\cite{Gold_PRB2014,Giesz_PRB2015,Thoma_APL2017,Reindl_NL2017} due to the variation in emission energy from dot to dot. Apart from selecting two quantum dots that happen to have the same emission energy, attempts have been made to dynamically tune two dots into resonance using controls such as applied strain\cite{Flagg_PRL2010} or electric-field \cite{Patel_NP2010}. An alternative to dynamic tuning is quantum dot intermixing which has been shown to be an effective method to tune emission energies over a wide spectral range\cite{Haysom_JAP2000,Haysom_SSC2000,Dubowski_APL2000,Girard_APL2004,Dion_APL2006,Dion_JAP2008,Dion_JAP2008b}. Intermixing is typically performed \textit{ex-situ} by rapid thermal annealing\cite{Girard_APL2004} but can also be done \textit{in-situ} using a focused laser\cite{Dubowski_APL2000}. $In$-$situ$ tuning with a focused laser has been demonstrated using quantum dot ensembles\cite{Rastelli_APL2007}.

In this study we investigate laser-induced intermixing processes in single InAsP/InP nanowire quantum dots. An InAsP/InP quantum dot is an inherently metastable system where if enough energy is provided, typically through heating, the As and P atoms will intermix through a diffusion process. This intermixing can be enhanced at lower temperatures by the presence of mobile crystal defects that are either already present (e.g. through judicious choice of crystal growth conditions\cite{Dion_APL2006} or ion implantation\cite{Dion_JAP2008}) or are thermally created. As these defects diffuse through the InAsP/InP quantum dot they cause an atomic mixing on the group V sites and hence a change in the emission energy of the dot. No change is observed on the group III sites since indium is the only group III atom present in these structures. This process is highly temperature dependent, with the creation of defects (typically at the semiconductor surface) requiring higher temperatures than the defect diffusion.

In addition to changes in emission energy, quantum dot intermixing can also affect material quality and lead to improved indistinguishability of the emitted photons. Defect states are known to trap carriers, leading to spectral fluctuations of the dot emission as the charge state fluctuates and reducing the indistinguishability of sequentially emitted photons\cite{Reimer_PRB2016}. This is particularly relevant in InP-based nanowire devices, where growth temperatures required for vapor-liquid-solid (VLS) epitaxy are lower than in conventional growth, leading to a higher density of impurities and crystal defects (e.g. group V interstitials\cite{Dion_APL2006}). Given that the low temperature intermixing process relies on the diffusion of grown-in defects and considering that these defects will eventually be annihilated or trapped at sinks such as exposed semiconductor surfaces, we expect that following the intermixing process we will be left with a material of improved crystal quality.

\begin{figure}
\begin{center}
\includegraphics*[width=8.8cm,clip=true]{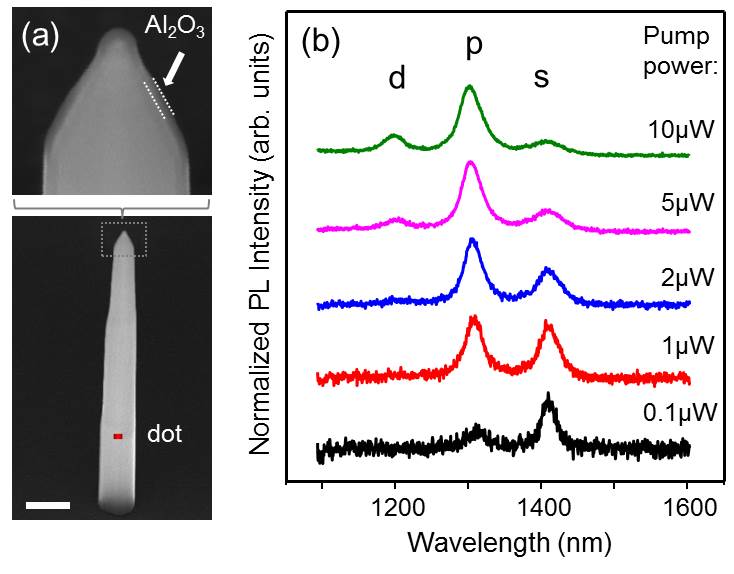}
\end{center}
\caption{(a) SEM image of a nanowire device imaged at $45^{\circ}$. The location of the quantum dot is schematically indicated by the red dot. Scale bar is 600\,nm. Upper panel is a higher magnification image of the nanowire tip showing the Al$_2$O$_3$ coating. (b) Room temperature power-dependent PL spectra from a single nanowire device containing a single quantum dot showing ground-state ($s$) and excited state ($p$, $d$) emission.}\label{fig1}
\end{figure}

The InAsP/InP nanowire quantum dot devices used in this study were grown using a combined selective-area and VLS epitaxy approach\cite{Dalacu_NT2009,Dalacu_NL2012} (see Methods). Importantly, the nanowires were grown using a position-controlled process\cite{Dalacu_NT2009} that ensures only a single nanowire within the intermixing laser spot, allowing for the tuning of individual quantum dots. To provide protection against surface decomposition during intermixing (see Supporting Information), the devices were coated with 40\,nm of Al$_2$O$_3$ or SiO$_2$, see Figure 1(a). Intermixing was carried out at a substrate temperatures of both 4\,K and 300\,K using a laser focused to a spot size of 2\,$\mu$m (see Methods). For the 300\,K measurements, devices emitting at telecom wavelengths\cite{Haffouz_NL2018} were used. The confining potentials in these dots are sufficiently deep to allow observation of ground and excited state emission at room temperature. For the 4\,K measurements, devices emitting around $\lambda = 930$\,nm were used.

Room temperature, power-dependent photoluminescence (PL) spectra from a typical telecom quantum dot used in the room temperature intermixing experiments are shown in Figure 1(b). At low pump powers, the ground-state emission is observed at $\lambda \sim 1400$\,nm. With increasing pump power, the atomic-like level structure is clearly evidenced by the appearance of peaks at higher energy which correspond to recombination from excited levels (e.g. $p$-shell and $d$-shell).

\begin{figure}
\begin{center}
\includegraphics*[width=7.5cm,clip=true]{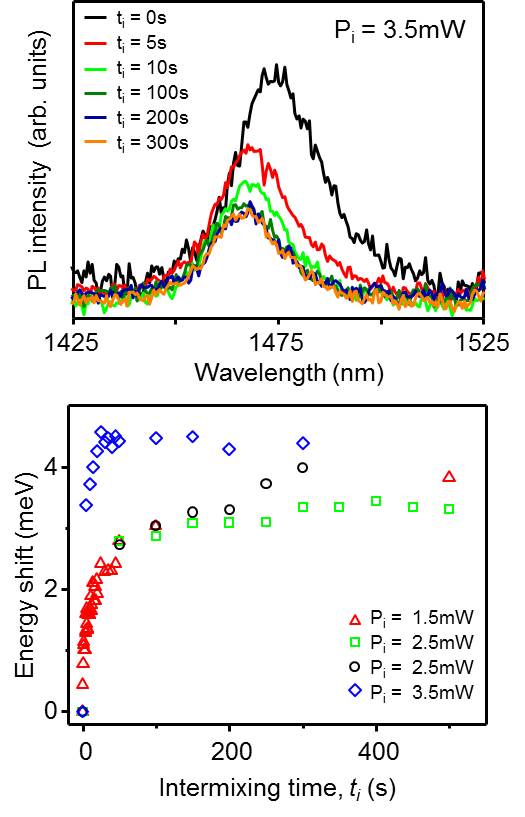}
\end{center}
\caption{(a) Room-temperature PL spectra from a single quantum dot at a function of intermixing time at an intermixing power $P_i=3.5$\,mW. (b) Time-dependence of the ground-state energy shift of four quantum dots intermixed at different powers.}\label{fig2}
\end{figure}

Ground-state emission wavelengths were determined from PL measurements at excitation powers $P=P_{sat}$, where $P_{sat} \sim 150$\,nW is the power required to saturate the ground-state transition. Intermixing was achieved by exciting at powers $P_{i}>10^4P_{sat}$ for various times $t_{i}$. The optical absorption of the pump laser in the nanowire resulted in a rapid heating to temperatures high enough to induce intermixing\cite{Dion_APL2006} (typically $T\sim 600^{\circ}$C). Figure 2(a) shows the 300\,K PL from the ground state of a single nanowire dot as a function of $t_{i}$ for an intermixing laser power of $3.5$\,mW using a 633\,nm laser. In Figure 2(b), we show the dependence on intermixing time of the ground-state energy shift for four nanowire devices annealed at different powers. All devices show a blue-shift of the ground state energy of approximately 3\,-\,4\,meV. The tuning behavior is non-linear,  with the rate decreasing and the energy shift finally saturating with increasing intermixing time. The sample exposed to the highest optical power exhibits the highest initial tuning rate. 

This behavior is typical of the intermixing process in InAsP/InP quantum dots with large numbers of grown-in P anti-site defects\cite{Dion_APL2006,Dion_JAP2008,Dion_JAP2008b}. Upon annealing, P anti-site defects dissociate to form P interstitials and In vacancies. As mentioned above, the highly mobile P interstitials subsequently diffuse through the quantum dot where the atomic intermixing of the group V species changes the composition profile of the dot, blue-shifting the emission energy as the InAsP dot material becomes increasingly P rich. This group V intermixing process can be described within a Fickian diffusion model\cite{Dion_JAP2008b}, using a temperature dependent inter-diffusion length $L_i \propto \sqrt{D_i t_i}$ where $D_i$ is the inter-diffusion constant. If the source of defects enabling this diffusion process is never depleted then $D_i$ is constant with time and the dot will intermix until its composition is very close to InP. In the nanowire case however, we observe a distinct saturation of the energy shift with time, consistent with the depletion of inter-diffusion-promoting defects in the system. We note that the largest shift obtained, ($> 4$\,meV) covers a significant fraction of the standard deviation of the emission energies typically observed in the nanowire system, $\sigma = 5.7$\,meV\cite{Chen_APL2016} allowing us to tune a large proportion of the nanowire dots to the same emission energy. 

\begin{figure}
\begin{center}
\includegraphics*[width=7.5cm,clip=true]{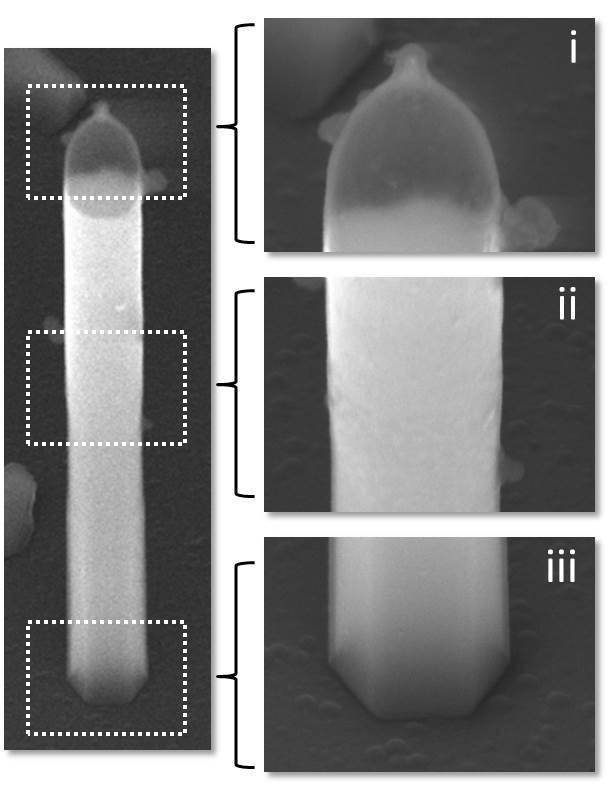}
\end{center}
\caption{SEM image of a nanowire after intermixing at $P_i = 3.5$\,mW for $t_i=500$\,s. Panels (i) through (iii) show melted, reflowed and intact sections of the nanowire respectively.}\label{fig3}
\end{figure}

As well as an energy shift, the intermixing process in the nanowire devices also results in a large reduction in the PL intensity, clearly seen in Figure 2(a). The integrated intensities of the four samples in Figure 2(b) decrease by a factor of four after the first few seconds of intermixing. The decease in brightness can be understood by looking at the nanowire after annealing. A scanning electron microscopy (SEM) image of a nanowire that has been annealed is shown in Figure 3, where significant differences from the as-grown devices are observed. We identify three different regions on the nanowire: (i) the top, where the InP has melted and left behind an inflated Al$_2$O$_3$ shell complete with a Au catalyst nipple, (ii) the middle, where melted InP from above is trapped underneath the Al$_2$O$_3$ shell and (iii) the bottom, which is intact, complete with clearly observable crystallographic sidewall facets.    

In Reference \citenum{Rastelli_APL2007}, the authors employed microdisk structures in which the pedestal limited the heat flow to the substrate. In such structures, the microdisk attains a uniform temperature which, for a given laser power, is determined by the thermal resistance of the pedestal. In the nanowire case, the temperature $T_L$ at a height $L$ above the substrate is expected to vary linearly according to (see Supporting Information):

\begin{equation}
T_L - T_{sub} \propto \frac{ P_{i} L}{ D_{nw}^2 K_o} 
\end{equation}where $T_{sub}$ is the temperature of the substrate, $D_{nw}$ is the nanowire diameter, and $K_0 = 0.068$\,mW\,$\mu$m$^{-1}$\,$^{\circ}$C$^{-1}$ is the thermal conductivity of InP. Clearly, exciting with sufficiently high powers to attain intermixing temperatures (i.e. $T\sim 600^{\circ}$C) at a dot height of $L\sim 1.5\,\mu$m results in temperatures exceeding the melting point of InP, $T_m = 1060^{\circ}$C, at the top of the nanowire, $L\sim 5\,\mu$m. The deterioration of the nanowire observed in Figure 3 will clearly modify the propagation of the waveguide mode (HE$_{11}$) which is crucial to obtaining high collection efficiencies in these devices\cite{Bulgarini_NL2014,Friedler_OE2009}.

\begin{figure}
\begin{center}
\includegraphics*[width=7.5cm,clip=true]{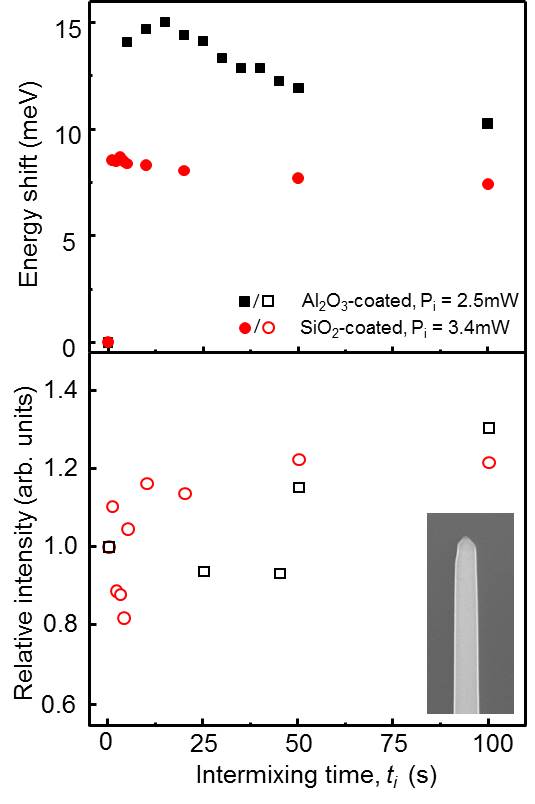}
\end{center}
\caption{Upper panel: energy shift of the ground state emission for devices annealed with the substrate held at 300K where the quantum dot is placed higher in the nanowire ($L=3.5\,\mu$m). Both Al$_2$O$_3$ and SiO$_2$ coated devices are shown. Lower panel: integrated intensity as a function of intermixing time normalized to pre-anneal values. Inset shows an SEM image of the top of the Al$_2$O$_3$ coated device after annealing.}\label{fig4}
\end{figure}

According to Equation 1, one can avoid melting the nanowire top while maintaining sufficiently high temperatures at the dot location simply by placing the dot higher in the nanowire and decreasing the laser power. Samples were grown with the quantum dot positioned $3.5\,\mu$m from the substrate (i.e. $2\,\mu$m higher than the devices shown in Figure 1, see Supporting Information). For these measurements, devices were prepared with either an Al$_2$O$_3$ or SiO$_2$ protective coating. The energy shift and integrated intensity as a function of intermixing time are shown in Figure 4 for both types of devices. The drop in intensity observed previously is absent, in fact these devices show a modest ($\sim 20\%$) increase in integrated intensity after annealing. SEM images taken after intermixing (see inset, Figure 4) show no signs of nanowire melting, consistent with this behavior. 

Due to the non-linear growth rate of these nanowire devices\cite{Dalacu_NT2009}, the additional core material grown to shift the quantum dots higher produces taller nanowires (7\,-\,8\,$\mu$m compared to 5\,-\,6\,$\mu$m) with a more pronounced taper. This being the case, we expect the energy shift of these Al$_2$O$_3$ coated devices to be different to that obtained with the shorter wires. In fact, with more defect-containing material above the quantum dot, one would expect a larger shift, as observed. We do not expect Al$_2$O$_3$ and SiO$_2$ coated devices to show the same shift considering the changes to the strain environment arising from the different coatings and the role that strain plays in determining the diffusion of defects in the system\cite{Barik_APL2007}. We also note that the power required to intermix these samples is similar to the powers which previously produced melting. The different geometry of the nanowires, and particularly the different taper, will modify the absorption profile in the wire\cite{Friedler_OE2009,Frederiksen_ACSP2017} and produce different temperatures in the nanowire for the same excitation power. 

For extended anneal times both samples in Figure 4(a) exhibit a slow red-shift of the ground-state, especially pronounced for the Al$_2$O$_3$ coated sample. The origin of this shift is unclear but may be related to coating-mediated changes in the strain-environment. Dielectric coatings on nanowires are known to shift quantum dot emission energies due to the strain that they apply\cite{Bavinck_NL2012}. For the coatings used in this study we observe a 22\,nm (14\,nm) red-shift (blue-shift) in the dot emission after SiO$_2$ (Al$_2$O$_3$) deposition indicating that SiO$_2$ (Al$_2$O$_3$) applies tensile (compressive) strain on the InP nanowire (see Supporting Information). To quantify the contribution of the observed energy shift with anneal due to a  coating-mediated change in the strain environment, we have remeasured the Al$_2$O$_3$-coated nanowires post-anneal with the Al$_2$O$_3$ removed using an HF wet-etch. If the annealing process does not change the strain environment, we expect a red-shift of 14\,nm. We observe slight red-shifts and even blue-shifts depending on the extent of the anneal (see Supporting Information).  This behavior suggests that the strain on the InP due to the Al$_2$O$_3$ coating transitions from compressive to tensile during the annealing process and is consistent with the red-shift observed for extended anneals. Evidence supporting structural changes in the Al$_2$O$_3$ with anneal based on wet-etch rates are given in Supporting Information. 

\begin{figure}
\begin{center}
\includegraphics*[width=7.5cm,clip=true]{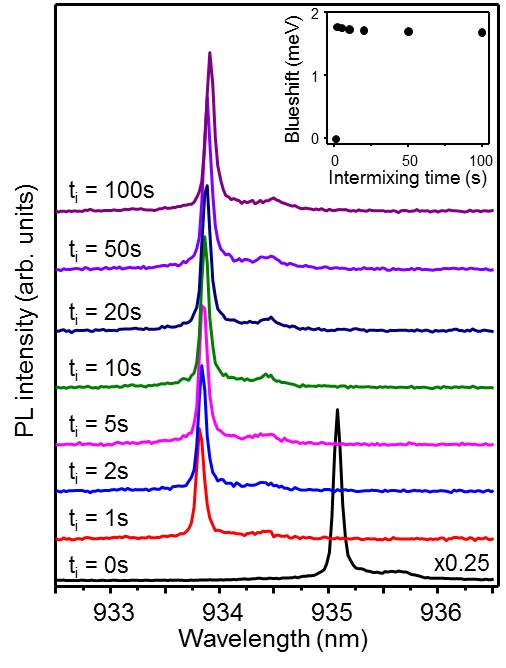}
\end{center}
\caption{Low-temperature PL spectra as a function of intermixing time using pulsed $P_i=3$\,mW. The inset shows the energy shift of the ground state with intermixing time.}\label{fig5}
\end{figure}

Finally, we have verified that the laser-induced intermixing approach also operates at low temperature (i.e. $T_{sub}=4$\,K). For these measurements, quantum dots emitting at $\lambda \sim 930$\,nm and positioned $L \sim 1.5\,\mu$m from the substrate were used and the protective coating was SiO$_2$. Irradiating at the maximum HeNe laser power available (3.5\,mW) did not result in any observable energy shift. To achieve high enough temperatures to promote intermixing, these samples were irradiated with a picosecond pulsed source at $\lambda = 800$\,nm with $P_i$ in the range 3 to 9\,mW.

PL spectra as a function of anneal time are shown in Figure 5. The intermixing shift is instantaneous on the scale of the shutter speed (500\,ms) and shows a time dependence (inset) very similar to the SiO$_2$ coated sample in Figure 4. The maximum shift obtained is $\sim 2$\,meV, substantially reduced compared to the sample in Figure 4. We attribute this reduction, as above, to differences in the nanowire geometries. In particular, the nanowire diameter in this device was tailored for emission at $\lambda\sim 930$\,nm (see Methods) and is much smaller than devices targeting telecom wavelengths (i.e. $D_{nw} = 250$\,nm compared to $D_{nw}=350$\,nm). Consequently, the volume of defect-containing InP material above the quantum dot is significantly reduced, resulting in a smaller energy shift. 

In conclusion, we have demonstrated an $in$-$situ$ technique to tune individual site-selected quantum dots using laser-induced intermixing. The observed shifts exceed the inhomogeneous broadening in our nanowire devices, thus offering a route to tune two arbitrary dots into resonance with a precision controlled through exposure time and limited by the rate of heat transfer to the substrate. We attribute the distinct saturation of the energy shift as indicative of improved material quality which has potential implications with regard to the ability to generate indistinguishable photons.  

\textbf{Sample Growth} The InAsP/InP nanowire devices were grown using chemical beam epitaxy on patterned Fe-doped InP (111)B substrates using a combined selective-area and VLS epitaxy approach\cite{Dalacu_NT2009}. The patterned substrates consisted of Au catalysts in the center of circular openings in a SiO$_2$ mask. A two stage growth was used in which an InP nanowire core with an embedded InAsP quantum dot was grown first using growth conditions that promote VLS (i.e. axial) growth. The nanowire core was subsequently clad with an InP shell using growth conditions that promote substrate and radial growth. The shell defines the photonic waveguide necessary for the efficient collection of the quantum dot emission\cite{Friedler_OE2009}. The diameter of the waveguide is dictated by the size of the SiO$_2$ opening and is tailored to the dot emission wavelength (i.e. $D_{nw}/\lambda \sim 0.23$)\cite{Haffouz_NL2018}. To provide protection against surface decomposition the devices were coated with 40\,nm of Al$_2$O$_3$ or SiO$_2$ deposited by atomic layer deposition (ALD) or plasma-enhanced chemical vapor deposition PECVD, respectively.  For details on the device fabrication and growth, see Refs. \citenum{Dalacu_NT2009}, \citenum{Dalacu_NL2012}, and \citenum{Haffouz_NL2018}

\textbf{Optical Spectroscopy.} Photoluminescence measurements on individual nanowires were performed with the wires still attached to the (111)B InP substrate. The 4\,K measurements were done in a continuous flow helium cryostat using above bandgap excitation (HeNe or Ti-Sapphire laser) through a 50X microscope objective (N.A. = 0.42) with a $\sim\,2\,\mu$m spot size. The long (short) wavelength PL was collected through the same microscope objective, dispersed using a 0.32\,m (0.5\,m) grating spectrometer and detected using a liquid-nitrogen cooled InGaAs diode (CCD) array. 





\providecommand{\latin}[1]{#1}
\makeatletter
\providecommand{\doi}
  {\begingroup\let\do\@makeother\dospecials
  \catcode`\{=1 \catcode`\}=2\doi@aux}
\providecommand{\doi@aux}[1]{\endgroup\texttt{#1}}
\makeatother
\providecommand*\mcitethebibliography{\thebibliography}
\csname @ifundefined\endcsname{endmcitethebibliography}
  {\let\endmcitethebibliography\endthebibliography}{}

\end{document}



\newpage
\section{Supplementary Material}

\subsection{Annealing with and without the protective dielectric coating}

All devices were coated with a thin dielectric layer to prevent material decomposition during the annealing process. The upper panel of Figure S1 illustrates the damage caused to the nanowires without the protective coating as the intermixing laser power is increased. We note that the damage is restricted to the top of the nanowires indicating that this is where absorption takes place. In the lower panel of Figure S1 we show the evolution of the structural changes that occur in coated nanowires. As the laser power is increased, the hollow section of the nanowire extends farther down and the top becomes inflated from the pressure exerted by released phosphorus. The pressure build up eventually causes the top of the coating to rupture (right image in the lower panel of Figure S1). 

\vspace{0.5cm}

\begin{figure}
\begin{center}
\includegraphics*[width=14cm,clip=true]{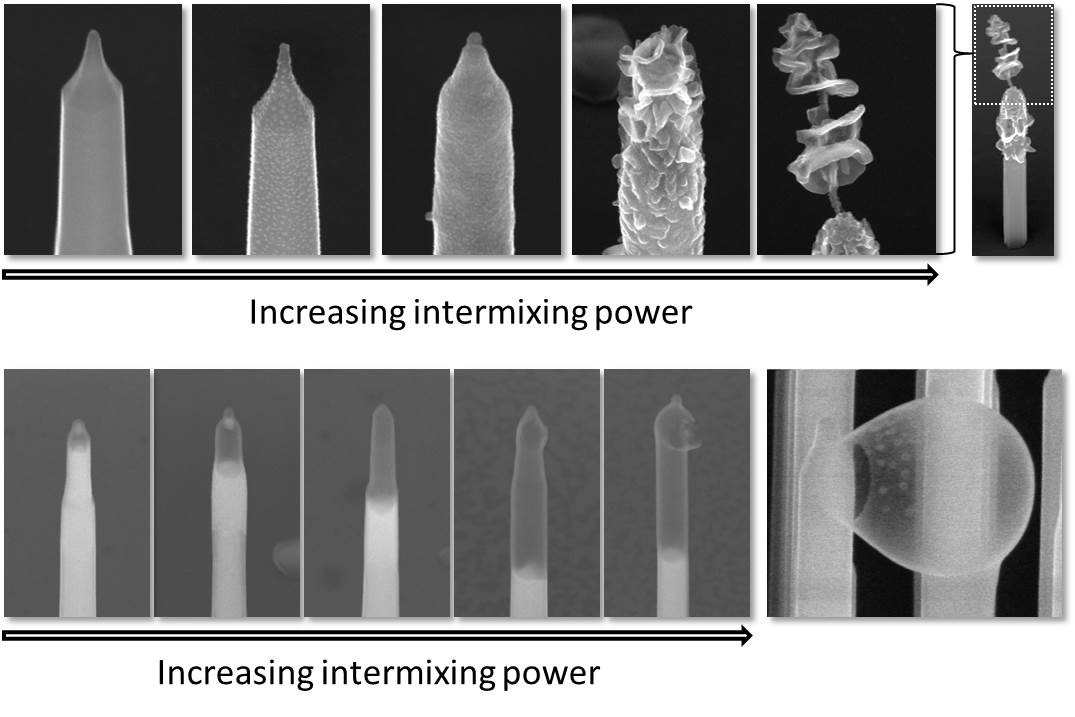}
\end{center}
\caption{Upper panel: SEM images of uncoated nanowires subjected to different laser powers ($< 2$\,mW). Lower panel: SEM images of nanowires coated with 40\,nm of Al$_2$O$_3$ for increasing laser powers ($> 2$\,mW).}\label{fig_s1}
\end{figure}
\newpage

\subsection{Temperature gradient in the nanowire}

If we assume that the incident laser power is absorbed at the nanowire tip, we can estimate the temperature distribution in the nanowire from the overlap of a Gaussian beam of diameter $D_{beam}$ with a cylinder of diameter $D_{nw}$ by solving a 1-dimensional heat conduction problem,  

\begin{equation}
T_L - T_{sub} = \frac{4P_{i} L}{\pi D_{nw}^2 K_o} \left[1 - exp\left[-\left(\frac{D_{nw}}{D_{beam}}\right)^2\right]\right]
\end{equation}where $T_{sub}$ is the temperature of the substrate and $K_0 = 0.068$\,mW\,$\mu$m$^{-1}$\,$^{\circ}$C$^{-1}$ is the thermal conductivity of InP. Figure S2 shows the temperature along the length of a nanwire for different incident powers, assuming all the light is absorbed (i.e. $D_{beam}=D_{nw}$). Although this simple model underestimates the temperature in the nanowire for a given incident power, it serves to illustrate that in order to attain temperatures required for intermixing\cite{Dion_APL2007} ($T\sim 600^{\circ}$C) at a dot height of $L\sim 1.5\,\mu$m,  temperatures at the top of the nanowire ($L\sim 5\,\mu$m) are likely to exceed the melting point of InP, $T_m = 1060^{\circ}$C.

\vspace{0.5cm}

\begin{figure}
\begin{center}
\includegraphics*[width=10cm,clip=true]{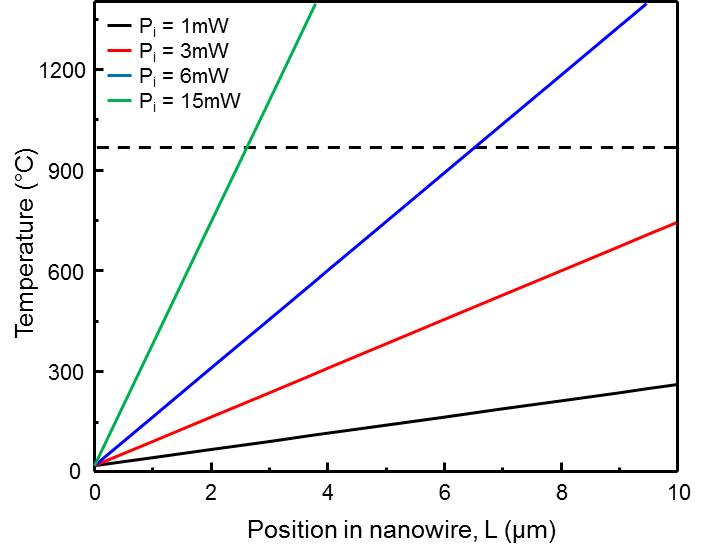}
\end{center}
\caption{Temperature of a nanowire along its length according to Equation 1 for different excitation powers. Dashed line indicates the melting temperature of InP.}\label{fig_s2}
\end{figure}

\newpage

\subsection{Control of the quantum dot position in the nanowire}

To shift the quantum dot position within the nanowire, the growth time of the nanowire core prior to incorporating the dot was increased from 15 to 25 minutes. This results in an increase of the height of the nanowire core (i.e. the position of the quantum dot within the nanowire waveguide) from $1.5\,\mu$m to $3.5\,\mu$m, as shown in Figure S3. The subsequent cladding growth is the same for both samples, which results in taller nanowires (7\,-\,$8\,\mu$m compared to 5\,-\,$6\,\mu$m) with longer tapers for the taller cores (see Figure S3).

\vspace{0.5cm}

\begin{figure}
\begin{center}
\includegraphics*[width=11cm,clip=true]{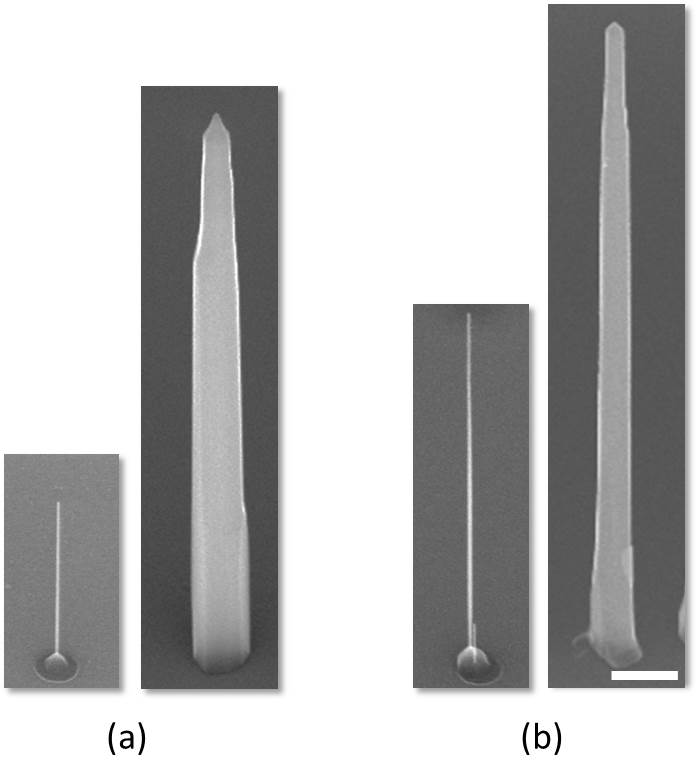}
\end{center}
\caption{(a) SEM images viewed at $45^{\circ}$ of a nanowire device grown using a $1.5\,\mu$m core before (left panel) and after (right panel) cladding. (b) Same but using a $3.5\,\mu$m nanowire core. Scale bar is 500\,nm.}\label{fig_s3}
\end{figure} 
 
\subsection{Influence of dielectric cladding on emission energy}

Dielectric coatings deposited on nanowires shift emission energies due to the strain that they apply on the semiconductor material\cite{Bavinck_NL2012}. The magnitude and direction of these shifts are dictated by the difference in thermal expansion coefficients between the dielectric coatings and nanowire material, and the temperature at which the dielectric coatings are deposited. For the Al$_2$O$_3$ coating used in this study we observe a 14\,nm blue-shift in the dot emission after deposition  (see Figure S4(a)) indicating that Al$_2$O$_3$ applies compressive strain on the InP nanowire. The SiO$_2$ coating, on the other hand, red-shifts the dot emission by 22\,nm (see Figure S4(b)) indicating that SiO$_2$ applies tensile strain on the InP nanowire.

\vspace{0.5cm}

\begin{figure}
\begin{center}
\includegraphics*[width=15cm,clip=true]{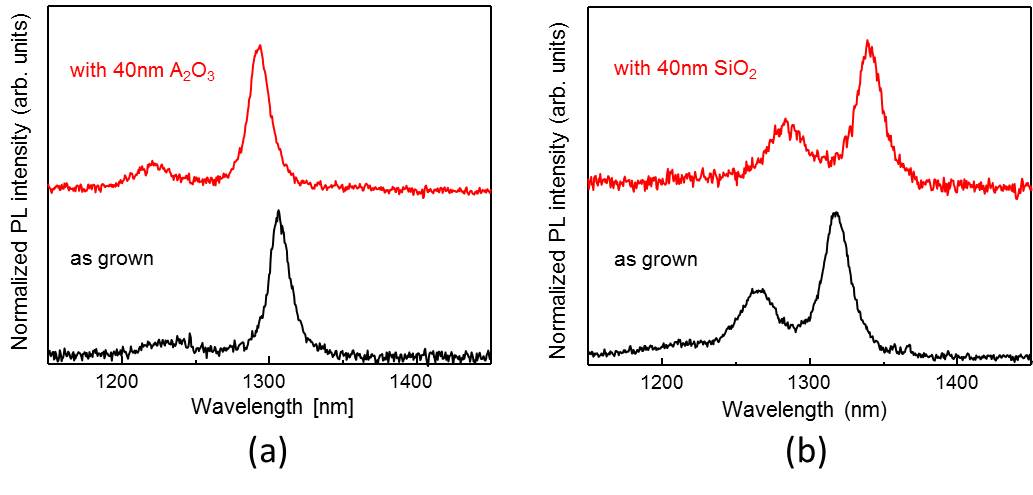}
\end{center}
\caption{Room temperature emission from two nanowire quantum dots as-grown and after deposition of 40\,nm of (a) Al$_2$O$_3$ and (b) SiO$_2$.}\label{fig_s4}
\end{figure}  

This process is completely reproducible. If, for example, the Al$_2$O$_3$ is subsequently removed by wet-etching in a dilute HF solution, the quantum dot emission energy red-shifts back to the as-grown value, as shown in Figure S5(a). If, however, the nanowire is annealed, stripping the Al$_2$O$_3$ produces a smaller red-shift, the magnitude depending on the intermixing laser power $P_i$ and the intermixing time $t_i$, and may even blue-shift slightly, as in the device shown in Figure 5(b).

\vspace{0.5cm}

\begin{figure}
\begin{center}
\includegraphics*[width=15cm,clip=true]{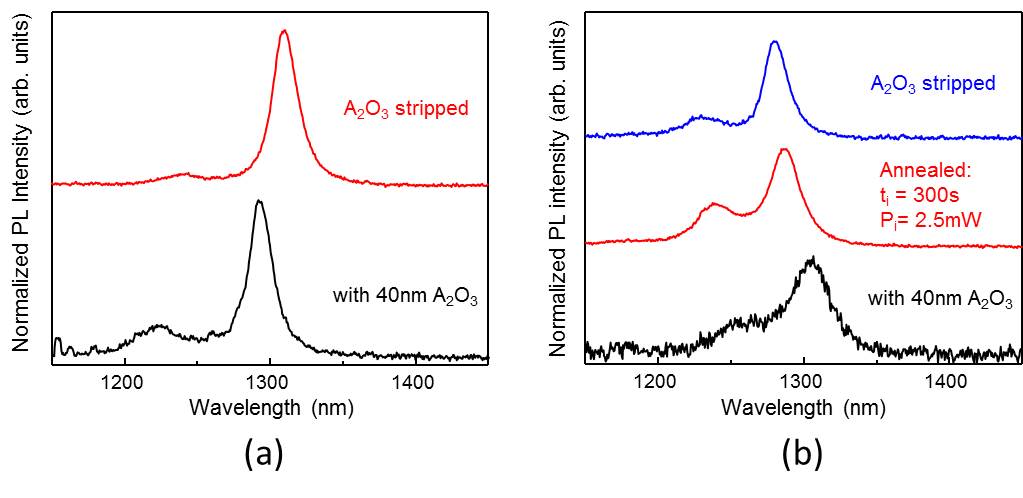}
\end{center}
\caption{(a) Room temperature emission from an unannealed quantum dot with the 40\,nm Al$_2$O$_3$ coating and after removal of the coating using an HF wet etch. (b) Same but for an annealed quantum dot.}\label{fig_s5}
\end{figure}  
 
We have used SEM to verify that the HF dip completely removes the Al$_2$O$_3$ coating. This is indeed the case for unannealed nanowires, as shown in Figure S6(a). For annealed nanowires, however, the HF dip only removes the Al$_2$O$_3$ from the bottom of the nanowires, leaving the coating on the tip intact, as shown in Figure S6(b).  The low temperature (200$^\circ$C) ALD process used to deposit the Al$_2$O$_3$ coatings produces films having an amorphous structure\cite{Prokes_APLM2014,Katz_JAP2016}. These films readily etch in dilute HF solutions, whereas the crystalline forms (e.g. corundum) do not.  The etching experiments suggest that the Al$_2$O$_3$ on the top part of the nanowire has undergone a crystallization to one of its crystalline forms. Similar crystallization has been observed in ALD thin films subjected to rapid thermal processing starting at temperatures of $T_c = 900^\circ$C\cite{Jakschik_TSF2003,Zhang_JPD2007}.  Assuming a linear temperature gradient in the nanowire with endpoints set by T$_{sub}$ and T$_m$, we determine a similar temperature threshold for crystallization from the extent of the unetched section of Al$_2$O$_3$ shown in Figure S6(b) (see Figure S6(c)).  

In conclusion, there is strong evidence of structural changes in the Al$_2$O$_3$ coating with annealing. The development of long range order that develops in the coating is likely to modify the strain environment of the quantum dot, and this would likely occur on a slower time scale compared to intermixing-mediated compositional changes. Such structural changes can explain the observed red-shift of the quantum dot emission for extended annealing as well as the energy shifts obtained upon removing the Al$_2$O$_3$ coating from annealed nanowires.

\vspace{0.5cm}

\begin{figure}
\begin{center}
\includegraphics*[width=14cm,clip=true]{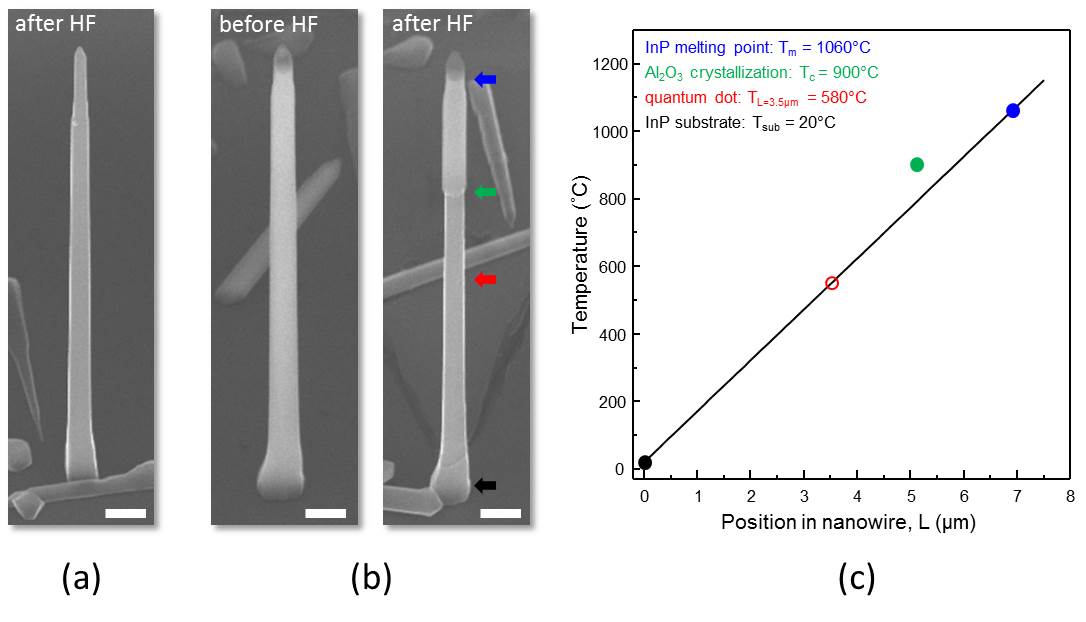}
\end{center}
\caption{(a) SEM image viewed at 45$^\circ$ of an unannealed nanowire after an HF dip demonstrating complete removal of the Al$_2$O$_3$. (b) Images before (left panel) and after (right panel) an HF dip of an annealed nanowire. In this case, the top section of the Al$_2$O$_3$ did not etch. Scale bars are 500\,nm. (c) Estimate of the position-dependent temperature in the nanowire during annealing based on T$_{sub}$ (black circle) and T$_m$ (blue circle). Also included is the position at which crystallization is observed (green circle) assuming a threshold temperature for crystallization, T$_{c}$, from Reference \citenum{Jakschik_TSF2003}. The red open circle is the predicted temperature at the dot position for a 3.5\,$\mu$m core.}\label{fig_s6}
\end{figure}  
 

\providecommand{\latin}[1]{#1}
\makeatletter
\providecommand{\doi}
  {\begingroup\let\do\@makeother\dospecials
  \catcode`\{=1 \catcode`\}=2\doi@aux}
\providecommand{\doi@aux}[1]{\endgroup\texttt{#1}}
\makeatother
\providecommand*\mcitethebibliography{\thebibliography}
\csname @ifundefined\endcsname{endmcitethebibliography}
  {\let\endmcitethebibliography\endthebibliography}{}